\documentclass[preprint,amsmath,amssymb,aps]{revtex4-1}
\usepackage{graphicx}
\usepackage{dcolumn}
\usepackage{bm}
\begin{document}


\title{
Detecting long-range attraction\\ 
between migrating cells\\ 
based on p-value distributions
}
\author{Claus Metzner}
\email{claus.metzner@gmail.com}
\affiliation{
\small 
Biophysics Group, Friedrich-Alexander University Erlangen, Germany
}
\date{\today}

\begin{abstract}
Immune cells have evolved to recognize and eliminate pathogens, and
the efficiency of this process can be measured 
in a Petri dish. Yet, even if the cells are time-lapse recorded and tracked with high resolution, it is difficult to judge whether the immune cells find their targets by mere chance, or if they approach them in a goal-directed way, perhaps using remote sensing mechanisms such as chemotaxis. To answer this question, we assign to each step of an immune cell a 'p-value', {\em the probability that a move, at least as target-directed as observed, can be explained with target-independent migration behavior}. The resulting distribution of p-values is compared to the distribution of a reference system with randomized target positions. By using simulated data, based on various chemotactic search mechanisms, we demonstrate that our method can reliably distinguish between blind migration and target-directed 'hunting' behavior. 
\end{abstract}

\maketitle


\section*{Introduction}
The ability of immune cells to find and kill pathogens in the body is critical for survival \cite{janeway1996immunobiology, kumar2011pathogen}, and is increasingly used in new immunotherapies for cancer treatment \cite{schuster2006cancer, rosenberg2014decade, drake2014breathing}. At the same time, important insights about the interaction between immune and cancer cells are being gained from highly controlled in-vitro experiments, in which the migration trajectories of all individual cells, as well as their interactions after steric contact, can be directly observed and quantitatively evaluated. 

In vivo, chemotaxis \cite{eisenbach04} plays a vital role in recruiting motile immune cells to sites of infection or to malignant cancers. This recruitment of immune cells is often based on endogenous chemo-attractants, which are released by other host cells that are already present at the location where a pathogen has invaded the body. However, the fact that individual immune cells are also able to find and eliminate cancer cells in-vitro, without being assisted by any other components of the immune system, suggests that immune cells may be guided by chemical traces or other cues produced by the cancer cells themselves. 

In this work, we develop a new statistical method which tests whether the immune cells are somehow attracted towards their targets, or if they simply perform a random walk that occasionally leads to chance encounters with a target cell. For this purpose, we consider an in-vitro assay where immune and target cells are randomly mixed together within a suitable matrix. We assume that the cells are time-lapse recorded with sufficient temporal resolution. Applying automatic tracking methods to the video recordings provides the individual cell trajectories, which are in the following approximated as sequences of straight moves between the recorded discrete positions.

The basic idea behind our method is that the migration behavior of an immune cell changes in a characteristic way as soon as it starts to home in on a target: compared to the immune cell's 'normal' migration behavior, a purposeful target approach will reveal itself by a larger probability of 'suspicious' moves that aim to align the immune cell's migration direction towards the target.

\begin{figure}[h!]
\centering
\includegraphics[width=17cm]{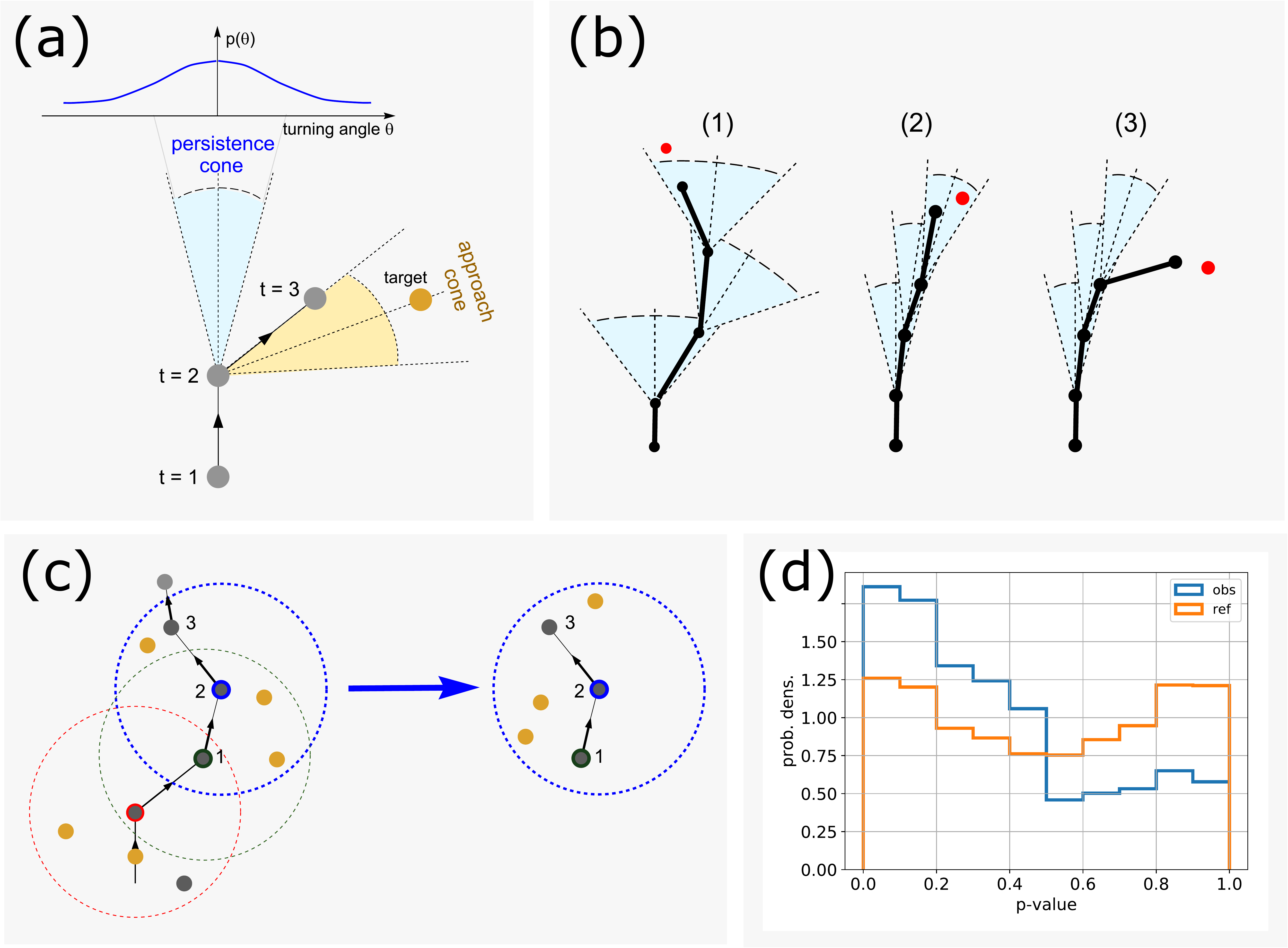}
\caption{
Explanation of our method to detect goal-directed migration. 
{\bf (a)} A 'triplet', consisting of three consecutive positions of a focal immune cell (gray circles), with a target cell (orange circle) in the vicinity. We consider the final move of the immune cell between time steps $t\!=\!2$ and $t\!=\!3$. The persistence cone (blue shaded area) is the interval of the immune cell's most probable migration directions, based on the known turning angle distribution $p(\theta)$. The approach cone (orange shaded area) is the interval of migration directions which are at least as target-oriented as the actual move of the immune cell. By integrating $p(\theta)$ over the approach cone, we compute a p-value, {\em the probability that a move at least as target-directed as observed could occur in a target-blind random walk}. A histogram of observed p-values is shown in (d).
{\bf (b)} Three examples of immune cell trajectories (black) in relation to a target cell (red). Cases (1) and (2) are not indicative of goal-directed migration, but case (3) is 'suspicious'.
{\bf (c)} A reference system without interactions between immune and target cells is generated by re-positioning the target cells randomly, while leaving the immune cell trajectory unchanged. A histogram of the resulting p-values is shown in (d).
{\bf (d)} Distribution of p-values in the observed data (blue line) and in the randomized reference system (orange line). The larger probability of small p-values in the observed data indicates that immune cells are attracted to the target cells.
\label{fig1}}
\end{figure}

Measuring this increased frequency of target-directed moves requires prior knowledge of the immune cell's 'normal' migration properties. We therefore first determine the probability distribution $p_i(\theta)$ of turning angles $\theta$ for each individual immune cell $i$. We then select out all those parts of the recorded cell trajectories where an immune cell is migrating in the vicinity of a potential target cell. More precisely, our analysis is based on {\em 'triplets'}: sequences of three successive video frames ($t=1,2,3$) in which the distance between a focal immune cell (the gray circles in Fig.~\ref{fig1}a) and some neighboring cancer cell (orange circle) is smaller than a pre-defined maximum interaction radius $r_{max}$. Within each triplet, we consider the final move of the immune cell between time steps $t\!=\!2$ and $t\!=\!3$.

Based on the turning angle distribution $p_i(\theta)$, we can define a {\em 'persistence cone'} as the interval of the immune cell's most probable migration directions (blue shaded area in Fig.~\ref{fig1}a). We can also define an {\em 'approach cone'} as the interval of migration directions which are at least as target-oriented as the actual move of the immune cell (orange shaded area). 

Finally, by integrating $p_i(\theta)$ over all turning angles $\theta$ within the approach cone, we can compute a {\em 'p-value'}, subsequently denoted by the symbol $\hat{p}$, and defined as {\em the probability that a move, at least as target-directed as observed, can be explained with target-independent migration behavior}. A very small p-value indicates a 'suspicious' move that provides a certain evidence for target-directed behavior.

Note that the p-value is small only if three conditions are simultaneously fulfilled: (A) The persistence cone is narrow, due to large directional persistence of the immune cell. (B) The approach cone is narrow, as the immune cell moves almost exactly towards the target cell. (C) The two cones are non-overlapping and distant from each other, because the immune cell is literally 'going out of its way' to approach the target. 

A typical suspicious move of the immune cell, corresponding to a small p-value, is shown in the last example (3) of Fig.~\ref{fig1}b. By contrast, a move isn't suspicious if one of the three conditions is missing. For instance, even if the immune cell is heading almost exactly towards the target, this can nevertheless be coincidental when the immune is generally moving with low directional persistence (Example (2) of Fig.~\ref{fig1}b). Reversely, for an immune cell with large directional persistence, approaching a target that is located well within the persistence cone is also not suspicious (Example (1) of Fig.~\ref{fig1}b). 

Finding just a few triplets with very low p-value does not provide convincing evidence for a target-directed immune cell migration in general. We therefore compute the distribution $prob_{obs}(\hat{p})$ of observed p-values over all evaluated triplets. The example distribution in Fig.~\ref{fig1}d (blue line) is based on simulated data, where the immune cells are able to home in on their targets by following spatial gradients of a chemo-attractant that is released by each target cell. 

Finally, we need to compare $prob_{obs}(\hat{p})$ with a reference distribution $prob_{ref}(\hat{p})$ of a system that resembles the observed one in all respects, except that there are no interactions between immune and cancer cells. To obtain this reference distribution, we use a bootstrapping method \cite{westfall1993resampling}: for each triplet, we leave the three positions of the immune cell unchanged, but shift all target cells that are located within the maximum interaction radius $r_{max}$ to new, independent random positions within that radius (Fig.~\ref{fig1}c). We then compute a histogram of the p-values based on these altered configurations to obtain $prob_{ref}(\hat{p})$ (orange line in Fig.~\ref{fig1}d). 

It turns out that the distribution of p-values in non-interacting systems is in general not uniform, nor does it correspond to any other standard distribution. However, any strong differences between the observed and the reference distribution, such as apparent in Fig.~\ref{fig1}d, indicate that the target cells somehow affect the migration of the immune cells. In particular, whenever the immune cells are attracted by the target cells, this manifests in a larger probability of small p-values in the observed distribution.


\section*{Results}

\subsection*{Validation of the method}

In order to validate our method, we apply it to artificial data from computer simulations of chemotactic behavior \cite{Metzner2019}. These simulations allow us to control the migration properties of the immune and target cells, and make it possible to switch between different chemotactic search strategies that might plausibly be used by actual immune cells.

\begin{figure}[h!]
\centering
\includegraphics[width=17cm]{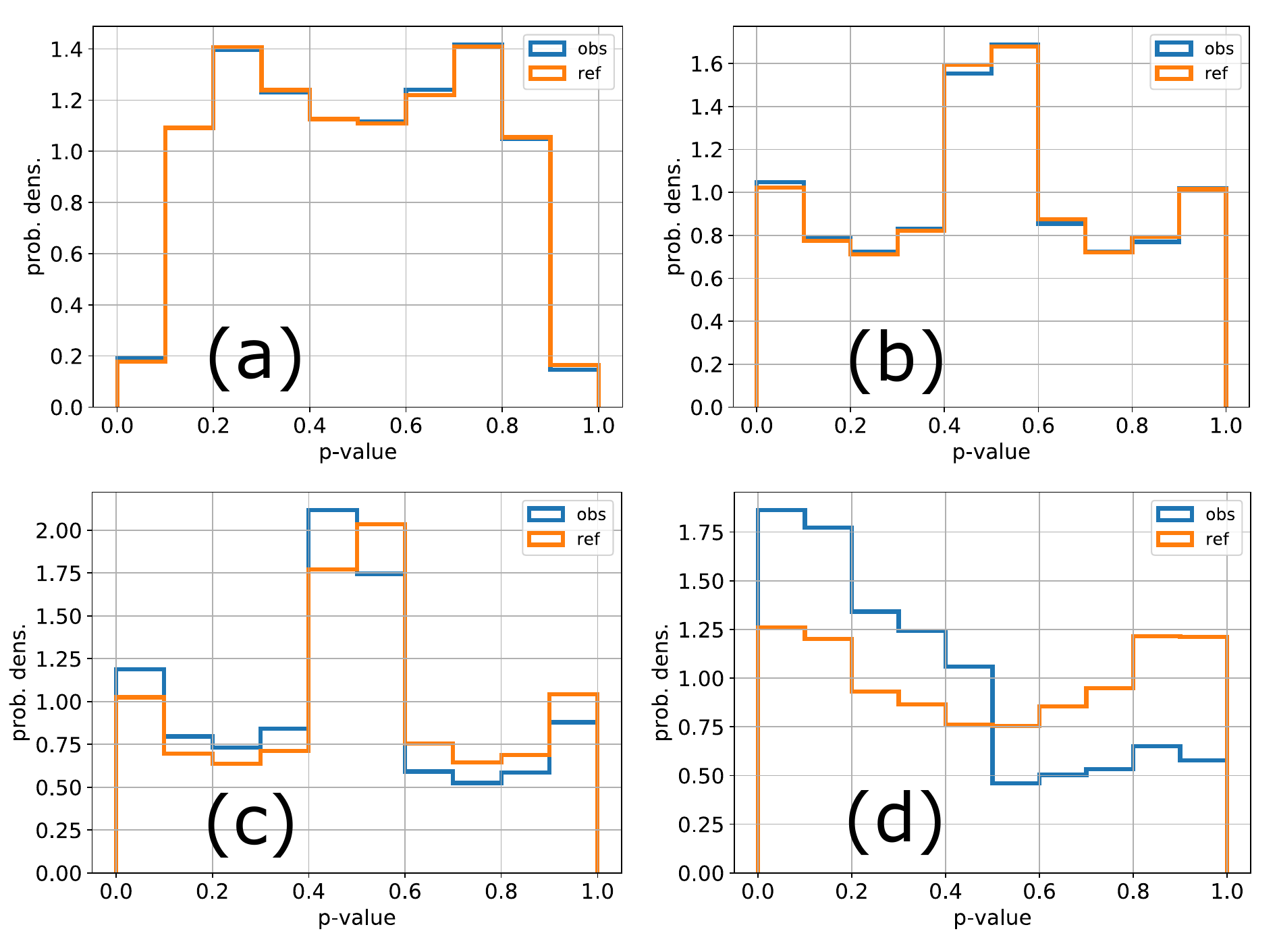}
\caption{
Application of our method to four different types of surrogate data. Shown are in each case the p-value distributions of the 'observed' system (blue lines) and of the randomized reference system (orange lines). In cases (a) and (b), there are no interactions between simulated immune and target cells, and the two distributions consequently coincide. In cases (c) and (d), chemotactic interactions between simulated immune and target cells lead to strong differences between the observed and reference distributions.
{\bf (a)} Target-blind, homogeneous migration: Simulated immune cells migrate according to a correlated random walk with temporally constant migration parameters.
{\bf (b)} Target-blind, heterogeneous migration: Simulated immune cells migrate according to a correlated random walk with temporally fluctuating migration parameters.
{\bf (c)} Temporal gradient sensing: Simulated immune cells use temporal gradients of a chemo-attractant to pursue the target cells.
{\bf (d)} Spatial gradient sensing: Simulated immune cells use spatial gradients of a chemo-attractant to pursue the target cells.
\label{fig2}}
\end{figure}

We start with a case where the immune cells do not interact at all with the targets but migrate 'blindly', according to a correlated random walk with fixed parameters for the mean step width (speed) and for the degree of directional persistence (For details, see 'Blind Search' in \cite{Metzner2019}). As expected, the resulting p-value distribution is identical to that of the randomized reference system (Fig.~\ref{fig2}(a)).

While cell migration can be well described as a correlated random walk with fixed parameters for short time scales (a few minutes), it has been demonstrated that migration parameters change gradually or abruptly on longer time scales \cite{Metzner2015}, even if the environment of the migrating cells is homogeneous, as on a plane Petri dish. If, accidentally, a change of migration parameters happens in the vicinity of a target cell, this may be miss-interpreted as a signal for long-range cell-cell interactions. To rule out this possibility, we next apply our method to a simulation in which the immune cells are still blind with respect to the targets, but occasionally switch between a highly persistent and a non-persistent (diffusive) migration mode (For details, see 'Random Mode Switching' in \cite{Metzner2019}). Although this heterogeneous type of migration changes the overall shape of the p-value distribution considerably, the observed and reference distributions are again identical (Fig.~\ref{fig2}(b)).

Next we turn to a case where the simulated immune cells actually approach the targets by following the temporal gradient of chemo-attractant (For details, see 'Temporal Gradient Sensing' in \cite{Metzner2019}). The used model assumes that the immune cells stay in a highly persistent migration mode as long as the concentration of chemo-attractant is increasing with time. When the concentration is decreasing, the immune cells switch to a diffusive mode in order to find a more goal-directed migration direction. Since this chemotactic mode switching resembles the random mode switching considered before, the overall shape of the p-value distribution is similar in Fig.~\ref{fig2}(c) and in Fig.~\ref{fig2}(b). Now, however, there are significant differences between the observed and reference distributions (blue and orange lines in Fig.~\ref{fig2}(c)). In particular, the observed distribution shows a larger probability of p-values smaller than 1/2, thus indicating attractive interactions. 

Finally, we consider a case where the simulated immune cells are able to sense the spatial gradient of chemo-attractant and to actively turn into the direction of a nearby target (For details, see 'Spatial Gradient Sensing' in \cite{Metzner2019}). This chemotactic approach strategy leads to yet another shape of the p-value distribution. More importantly, since target-directed turns of the immune cells are considered as highly 'suspicious' moves in our method, we now find very large differences between the observed and reference distributions (blue and orange lines in Fig.~\ref{fig2}(d)).


\section*{Methods}
\subsection*{Quasi-2D and 3D essays}

We assume an experimental assay where immune and cancer cells are mixed together in a collagen gel, or in any other matrix which is suitable for effective cell migration and which enables proper imaging with a microscope. If the matrix layer has a vertical thickness of only a few cell diameters, the system can be considered quasi two-dimensional, and the subsequent analysis can be restricted to the horizontal (x,y) cell positions. In the case of thicker matrices, where two cells can have the same horizontal position but be in different vertical planes, the z-position of the cells has to be measured as well, which is often not possible with very high precision For this reason, out method is strongly focused on the horizontal cell coordinates. The z-coordinates are only used to select pairs of immune and target cells from similar z-planes as possible interaction partners.

\subsection*{Format of input data}

We assume that the cells in a given field of view are time-lapse recorded with sufficient spatial and temporal resolution. Automatic tracking methods can then be used to extract from each video frame the momentary cell configuration, which is stored in a separate file for later convenience. Each configuration file should contain a list of lines in the form  $(x,y,z,i,c)$, with each line corresponding to a specific cell. Here, $x,y,z$ are the coordinates of the cell center, $i$ is an ID number that is unique to each cell and that persists over subsequent video frames, and $c\in\left\{0=immune,1=target\right\}$) is the category of the cell. The number of lines in the configuration files can change from one time point to the next, as cells may leave or enter the microscope's field of view, because of cell division and death, or due to tracking problems. 

\subsection*{2D cell migration model}

From the configuration files, we extract the temporal trajectory of each individual cell $i$, defined as the list of 3D positions $\vec{R}^{(i)}_t = (x^{(i)}_t,y^{(i)}_t,z^{(i)}_t)$ for successive time indices $t=0,1,2,\ldots$. For our migration model, we need only the 2D positions, denoted by $\vec{r}^{(i)}_t = (x^{(i)}_t,y^{(i)}_t)$. 

The sequence of a cell's horizontal positions $\vec{r}^{(i)}_t$ is approximated by a directionally persistent random walk with a certain distribution $p_i(w)$ of step widths $w$, and a distribution $p_i(\theta)$ of turning angles $\theta$. Here, the step width in the move from time $t$ to $t\!+\!1$ is defined as $w=|\vec{r}^{(i)}_{t\!+\!1}-\vec{r}^{(i)}_t|$, and the turning angle is defined as the angle between the two shift vectors $\left[\vec{r}^{(i)}_{t\!+\!1}-\vec{r}^{(i)}_t\right]$ and $\left[\vec{r}^{(i)}_t-\vec{r}^{(i)}_{t\!-\!1}\right]$.

The step width distribution is modeled as a Rayleigh distribution with speed parameter $\sigma_i$:
\begin{equation}
p_i(w) = \frac{w}{\sigma_i^2} \exp\left( -\frac{1}{2}\frac{w^2}{\sigma_i^2}\right).
\end{equation}

The turning angle distribution is modeled as a von Mises distribution with persistence parameter $\kappa_i$:
\begin{equation}
p_i(\theta) = \frac{1}{2\pi I_0(\kappa_i)} \exp\left( \kappa_i\cdot\cos(\theta) \right).
\end{equation}

Note that the speed and persistence parameters can be efficiently estimated from the time series of step widths and turning angles \cite{evans2000statistical}. The two parameters $\sigma_i$ and $\kappa_i$ describe the 'normal' (average) migration properties of each individual cell $i$.

\subsection*{Triplet-based analysis}

After the determination of the 'normal' cell migration properties, our method analyzes the motion of the individual immune cells in the context of their surrounding target cells. It is of practical importance that the cells need not to be tracked consecutively over a large number of frames, as our method requires only short 'triplets': sequences of three successive frames in which the positions of the same immune cell $i$ and of at least one nearby target cell $j$ (located within a three-dimensional sphere of radius $r_{max}$) are available. If a cell trajectory contains tracking gaps, the specific triplets containing such gaps are excluded from the analysis, but all other triplets are being used.

\subsection*{Observed p-values}

From each triplet we obtain three successive positions 
$\vec{r}^{(i)}_{t\!-\!1}$, $\vec{r}^{(i)}_t$, and $\vec{r}^{(i)}_{t\!+\!1}$ of immune cell $i$, as well as the position $\vec{r}^{(j)}_t$ of target $j$. The immune cell's shift vector $\vec{s}_1=\vec{r}^{(i)}_{t\!+\!1}-\vec{r}^{(i)}_t$ encloses a certain angle $|\phi|$ with the relative vector $\vec{u}_{ij}=\vec{r}^{(j)}_t-\vec{r}^{(i)}_t$ between immune and target cell. Note that there exists also another (hypothetical) shift vector $\vec{s}_2$ that encloses the same angle $|\phi|$ with the relative vector $\vec{u}_{ij}$. The range of directional angles $\left[\alpha_1,\alpha_2\right]$ enclosed by $\vec{s}_1$ and $\vec{s}_2$ is called the 'approach cone' (orange shaded area in Fig.~\ref{fig1}(a)). The interval $\left[\alpha_1,\alpha_2\right]$ of directional angles can be translated into an interval $\left[\theta_1,\theta_2\right]$ of turning angles for the immune cell. Choosing any turning angle in this interval would have aligned the immune cell with the target {\em at least as much} as in the immune cell's actual move. We can therefore compute a p-value as
\begin{equation}
\hat{p} = \int_{\theta_1}^{\theta_2} p_i(\theta) d\theta.
\end{equation}
If there is more than one target in the triplet, a separate p-value is computed for each target. The same procedure is repeated for all triplets of immune cell $i$, and for all other immune cells $i^{\prime}\neq i$ in the same way. All p-values are pooled, and a histogram finally yields the distribution $prob_{obs}(\hat{p})$.

\subsection*{Reference p-values} 
 
In order to obtain a reference distribution of p-values without any interactions between immune and target cells, we use a bootstrapping method \cite{westfall1993resampling}: for each triplet, we leave the three positions of the immune cell unchanged, but shift all target cells that are located within the maximum interaction radius $r_{max}$ to new, independent random positions within that radius (Fig.~\ref{fig1}c). We then compute a histogram of the p-values based on these altered configurations to obtain $prob_{ref}(\hat{p})$ (orange line in Fig.~\ref{fig1}d).


\section*{Discussion}
In this work, we have addressed the question of whether immune cells in a Petri dish find their targets by chance, or are attracted to the targets by some long-range interactions. This question has the form of a statistical hypothesis test, with the null hypothesis being that the immune cells perform a free random walk, independently of the target positions. Therefore, each step of an immune cell can be associated with a p-value, {\em the probability that a step at least as target-directed as observed could occur in a free (target-blind) random walk}. 

Recently, and for good reasons, the misuse of p-values has been strongly criticized in the scientific community \cite{gelman2006difference, goodman2008dirty, johnson2013revised, kyriacou2016enduring}. The core of the problem is that many research studies treat the p-value as a uniquely defined feature of their experiment, whereas there actually exists a (meta-) probability distribution for the p-value \cite{sackrowitz1999p, taleb2016short}: When the very same experiment is repeated (that is, when new samples are drawn from the very same statistical model), the p-value will fall sometimes below and sometimes above the significance level. Picking just a single p-value thereby leads to non-reproducible results.

For this reason, our method does not rely on a single p-value  relative to some arbitrary level of significance. Instead, we compute the complete distribution $prob_{obs}(\hat{p})$ of p-values, pooled over all recorded steps of the immune cells, and we compare the observed distribution with that of a randomized reference system $prob_{ref}(\hat{p})$. If there are long-range attractions between immune and target cells, small p-values will be more pronounced in $prob_{obs}(\hat{p})$ than in $prob_{ref}(\hat{p})$.

We have validated the method using simulated data, assuming two cases where the immune cells perform a free (target-blind) random walk, as well as two cases where the immune cells are using temporal or spatial chemo-attractant gradients to home in on the targets. Our method shows almost identical distributions $prob_{obs}(\hat{p})$ and $prob_{ref}(\hat{p})$ in the first two cases, but a strong enhancement of small p-values in the last two cases. We therefore conclude that the presented method can reliably distinguish between target-blind migration and purposeful pursuit.


\bibliographystyle{unsrt}
\bibliography{references}

\begin{thebibliography}{10}

\bibitem{janeway1996immunobiology}
Charles~A Janeway, Paul Travers, Mark Walport, Mark Shlomchik, et~al.
\newblock {\em Immunobiology: the immune system in health and disease},
  volume~7.
\newblock Current Biology London, 1996.

\bibitem{kumar2011pathogen}
Himanshu Kumar, Taro Kawai, and Shizuo Akira.
\newblock Pathogen recognition by the innate immune system.
\newblock {\em International reviews of immunology}, 30(1):16--34, 2011.

\bibitem{schuster2006cancer}
Manfred Schuster, Andreas Nechansky, and Ralf Kircheis.
\newblock Cancer immunotherapy.
\newblock {\em Biotechnology Journal: Healthcare Nutrition Technology},
  1(2):138--147, 2006.

\bibitem{rosenberg2014decade}
Steven~A Rosenberg.
\newblock Decade in review—cancer immunotherapy: entering the mainstream of
  cancer treatment.
\newblock {\em Nature Reviews Clinical Oncology}, 11(11):630, 2014.

\bibitem{drake2014breathing}
Charles~G Drake, Evan~J Lipson, and Julie~R Brahmer.
\newblock Breathing new life into immunotherapy: review of melanoma, lung and
  kidney cancer.
\newblock {\em Nature reviews Clinical oncology}, 11(1):24, 2014.

\bibitem{eisenbach04}
Michael Eisenbach.
\newblock {\em Chemotaxis}.
\newblock World Scientific Publishing Company, 2004.

\bibitem{westfall1993resampling}
Peter~H Westfall, S~Stanley Young, et~al.
\newblock {\em Resampling-based multiple testing: Examples and methods for
  p-value adjustment}, volume 279.
\newblock John Wiley \& Sons, 1993.

\bibitem{Metzner2019}
Claus Metzner.
\newblock {Principles of efficient chemotactic pursuit}.
\newblock {\em arXiv}, 1902.10589:1--27, 2019.

\bibitem{Metzner2015}
Claus Metzner, Christoph Mark, Julian Steinwachs, Lena Lautscham, Franz
  Stadler, and Ben Fabry.
\newblock {Superstatistical analysis and modelling of heterogeneous random
  walks.}
\newblock {\em Nature communications}, 6(May):7516, jun 2015.

\bibitem{evans2000statistical}
Merran Evans, Nicholas Hastings, and Brian Peacock.
\newblock Statistical distributions.
\newblock 2000.

\bibitem{gelman2006difference}
Andrew Gelman and Hal Stern.
\newblock The difference between significant and not significant is not itself
  statistically significant.
\newblock {\em The American Statistician}, 60(4):328--331, 2006.

\bibitem{goodman2008dirty}
Steven Goodman.
\newblock A dirty dozen: twelve p-value misconceptions.
\newblock In {\em Seminars in hematology}, volume~45, pages 135--140. Elsevier,
  2008.

\bibitem{johnson2013revised}
Valen~E Johnson.
\newblock Revised standards for statistical evidence.
\newblock {\em Proceedings of the National Academy of Sciences},
  110(48):19313--19317, 2013.

\bibitem{kyriacou2016enduring}
Demetrios~N Kyriacou.
\newblock The enduring evolution of the p value.
\newblock {\em Jama}, 315(11):1113--1115, 2016.

\bibitem{sackrowitz1999p}
Harold Sackrowitz and Ester Samuel-Cahn.
\newblock P values as random variables - expected p values.
\newblock {\em The American Statistician}, 53(4):326--331, 1999.

\bibitem{taleb2016short}
Nassim~Nicholas Taleb.
\newblock A short note on p-value hacking.
\newblock {\em arXiv preprint arXiv:1603.07532}, 2016.

\end{thebibliography}


\section*{Author contributions statement}
CM developed the concept, implemented the method, and wrote the paper.

\section*{Additional information}

\subsection*{Funding}
This work was funded by the Grant ME1260/11-1 of the German Research Foundation DFG.

\subsection*{Competing interests statement }
The authors declare no competing interests.

\subsection*{Data availability statement}

The chemotaxis simulation program (in C++, including videos) which was used to generate surrogate data is available at {\em http://tinyurl.com/cm-chemotactic-pursuit}.
The evaluation program based on the proposed method (in C++, including Python script for plotting and sample data) is available at {\em http://tinyurl.com/cm-pvaluemethod }.

\subsection*{Ethical approval and informed consent}
Not applicable.

\subsection*{Third party rights}
All material used in the paper are the intellectual property of the authors.


\clearpage

\end{document}